\documentclass[fleqn,twoside]{article}
\usepackage{espcrc2,epsf}

\newcommand{\AmS}{{\protect\the\textfont2
  A\kern-.1667em\lower.5ex\hbox{M}\kern-.125emS}}

\def\lsim{\raise0.3ex\hbox{$<$\kern-0.75em\raise-1.1ex\hbox{$\sim$}}}
\def\gsim{\raise0.3ex\hbox{$>$\kern-0.75em\raise-1.1ex\hbox{$\sim$}}}

\title{
Charmonium spectrum from quenched QCD on anisotropic lattices
\thanks{Talk presented by M.~Okamoto
}
}

\author{CP-PACS Collaboration : 
  S.~Aoki\rlap,\address{Institute of Physics,
    University of Tsukuba, Tsukuba, Ibaraki 305-8571, Japan}
  R.~Burkhalter\rlap,$^{\rm a,b}$
  S.~Ejiri\rlap,\address{Center for Computational Physics,
     University of Tsukuba, Tsukuba, Ibaraki 305-8577, Japan}
     \thanks{present address: Department of
  Physics, University of Wales Swansea, Singleton Park, Swansea SA2 8PP, UK}
  M.~Fukugita\rlap,\address{Institute for Cosmic Ray Research,
    University of Tokyo, Kashiwa 277-8582, Japan}
  S.~Hashimoto\rlap,\address{High Energy Accelerator Research Organization
    (KEK), Tsukuba, Ibaraki 305-0801, Japan}
  N.~Ishizuka\rlap,$^{\rm a,b}$
  Y.~Iwasaki\rlap,$^{\rm a,b}$
  K.~Kanaya\rlap,$^{\rm a}$
  T.~Kaneko\rlap,$^{\rm d}$
  Y.~Kuramashi\rlap,$^{\rm d}$
  V.~Lesk\rlap,$^{\rm b}$
  K.~Nagai\rlap,$^{\rm b}$\thanks{present address: Theory Division, CERN, 
  CH-1211 Geneva 23, Switzerland}
  M.~Okamoto\rlap,$^{\rm a}$
  M.~Okawa\rlap,$^{\rm d}$
  Y.~Taniguchi\rlap,$^{\rm a}$
  A.~Ukawa$^{\rm a,b}$ and
  T.~Yoshi\'e$^{\rm a,b}$
  }

\begin{document}

\begin{abstract}
 We present our final results of the charmonium spectrum 
 in quenched QCD on anisotropic lattices. Simulations are made 
 with the plaquette gauge action  and a tadpole improved clover 
 quark action employing $\xi = a_s/a_t = 3$. We calculate 
 the spectrum of S- and P-states and their excitation, 
 and study the scaling behavior of mass splittings. 
 Comparison is made with the experiment and previous lattice results.
 The issue of hyperfine splitting for different choices of the clover 
 coefficients obtained by Klassen is discussed.
\end{abstract}

\maketitle
\setcounter{footnote}{0}

\section{Introduction}

Standard lattice QCD actions on space-time isotropic 
lattices encounter serious obstacles for heavy quarks with
currently accessible lattice spacings 
because mass-dependent $O(ma)$ discretization errors
are very large. Aiming to reduce such errors,
Klassen\cite{Klassen98,Klassen} 
has proposed to employ anisoropic lattices
with $ma_t \ll 1$ for heavy quark simulations.
In this paper, we summarize our final results of the quenched charmonium
spectrum using the anisotropic method\cite{cppacs,cppacs2}.
We also address the problem with hyperfine splitting\cite{Klassen} 
that different choices of clover coefficients lead to disagreeing 
results in the continuum limit. 

\section{Simulations}

\begin{table} [t]
\setlength{\tabcolsep}{0.5pc}
\begin{center}
\caption{Simulation parameters. 
$a_s$ is fixed by $r_0=0.5$~fm.}
  \begin{tabular}{ccccc}
\hline
$\beta$ & $a_s^{r_0}$[fm]
&$L^3 \times T$  & $L a_s $[fm] & \#conf \\ \hline
5.7 &0.204 &$8^3  \times 48$ & 1.63 & 1000 \\
5.9 &0.137 &$12^3 \times 72$ & 1.65 & 1000\\
6.1 &0.099 &$16^3 \times 96$ & 1.59 & 600\\ 
6.35 &0.070 & $24^3 \times 144$ & 1.67   & 400\\
\hline
\end{tabular}
\vspace{-10mm}
\end{center}
\label{tab:param}
\end{table}

We use the standard anisotropic gauge action given by
$
S_g = \beta \sum (1/\xi_0 P_{ss'} + \xi_0 P_{st}).
$
The bare anisotropy $\xi_0$ is tuned to obtain a desired value of 
the renormalized anisotropy $\xi \equiv a_s/a_t$, adopting Klassen's
parametrization\cite{calib}. 

For quark we use an 
anisotropic clover quark action:
\begin{eqnarray}
S_f 
\hspace{-.5cm} 
&&=  \sum \{ \bar{\psi}_x \psi_x 
 \nonumber\\
&&\hspace{-.5cm} - K_t [ \bar{\psi}_x (1- \gamma_0) U_{0,x} \psi_{x+\hat{0}} 
+ \bar{\psi}_{x+\hat{0}} (1+\gamma_0) U^{\dagger}_{0,x} \psi_{x}] 
 \nonumber\\
&&\hspace{-.5cm}- K_s [ \bar{\psi}_x (1- \gamma_i) U_{i,x} \psi_{x+\hat{i}} 
+ \bar{\psi}_{x+\hat{i}} (1+\gamma_i) U^{\dagger}_{i,x} \psi_{x}] 
 \nonumber\\
&&\hspace{-.5cm}+ i K_s [c_s \bar{\psi}_x \sigma_{ij} F_{ij}(x) \psi_x
+ c_t \bar{\psi}_x \sigma_{0i} F_{0i}(x) \psi_x]\}.
\end{eqnarray}
The bare quark mass is given by $m_{0} = 1/2K_t - 3/\zeta -1$ with 
$\zeta \equiv K_t/K_s$.
For $\zeta$ we adopt the tree level 
tadpole improved value for massive quarks. 
For clover coefficients $c_s$ and $c_t$, we employ the values in 
the massless limit.
We note that our choice of $c_s$ is still correct for 
massive quarks because it has no mass dependence at the tree 
level\cite{cppacs2}.
The tadpole factors are determined as 
$\langle U_s \rangle = \langle P_{ss'} \rangle^{1/4}$ 
with $P_{ss'}$ the spatial plaquette
and $\langle U_t \rangle =1$.

Simulation parameters are summarized in Table~1.
We adopt lattices with $\xi =3$ and $La_s \sim 1.6$~fm. 
Runs are made at four values of $\beta$ which correspond to
$a_s=0.07$-0.20~fm.
For each $\beta$, we measure S- and P-state meson correlation functions
at two values of bare quark mass. Results are then inter(extra)polated to  
the charm quark mass where $1\bar{S}$ mass has its experimental value.
The lattice scale is set by either the Sommer scale $r_0=0.5$~fm,   
$1\bar{P} - 1\bar{S}$ splitting or $2\bar{S} - 1\bar{S}$ splitting.

\section{Results}

\begin{figure}[tb]
\vspace{-5mm}
\begin{center}
\leavevmode
\epsfxsize=6.7cm
\epsfbox{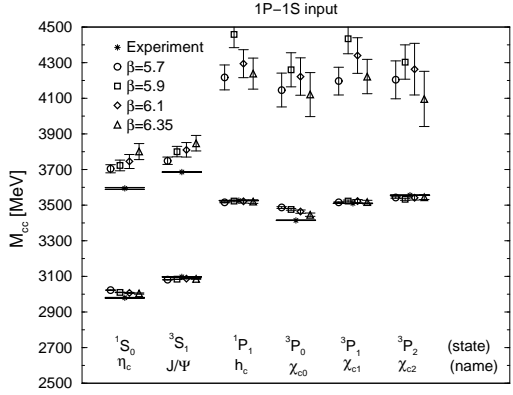}
\end{center}
\vspace{-8mm} 
    \caption{$c\bar{c}$ spectrum with $1\bar{P} - 1\bar{S}$ input.}
    \label{fig:mass}
\vspace{-15pt}
\end{figure}

In Fig.\ref{fig:mass}, we show results of the charmonium spectrum with the scale
from the $1\bar{P} - 1\bar{S}$ splitting.
Gross features of the spectrum are consistent with the experiment, {\it e.g.}
splittings between $\chi_c$ states are well resolved with correct
ordering. The deviation of 2S masses from the experiment is in part ascribed
to the quenching effect and in part to contaminations from higher
excited states.

\subsection{Hyperfine splitting}

\begin{figure}[t]
\vspace{-1.5mm}
\begin{center}
\leavevmode
\epsfxsize=6.7cm
\epsfbox{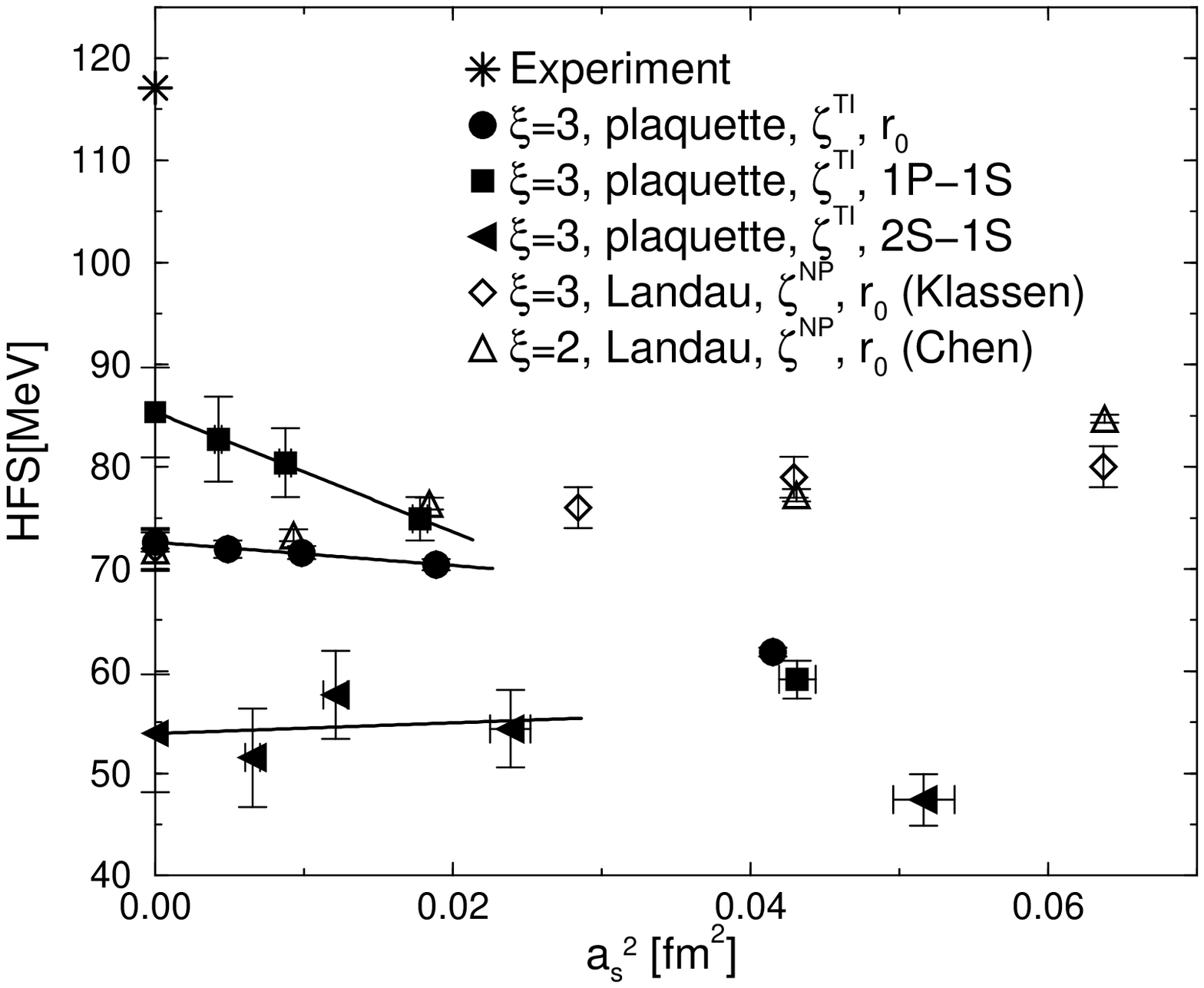}
\end{center}
\vspace{-16mm}
    \caption{Hyperfine splitting.}
    \label{fig:hfs}
\begin{center}
\leavevmode
\epsfxsize=6.5cm
\epsfbox{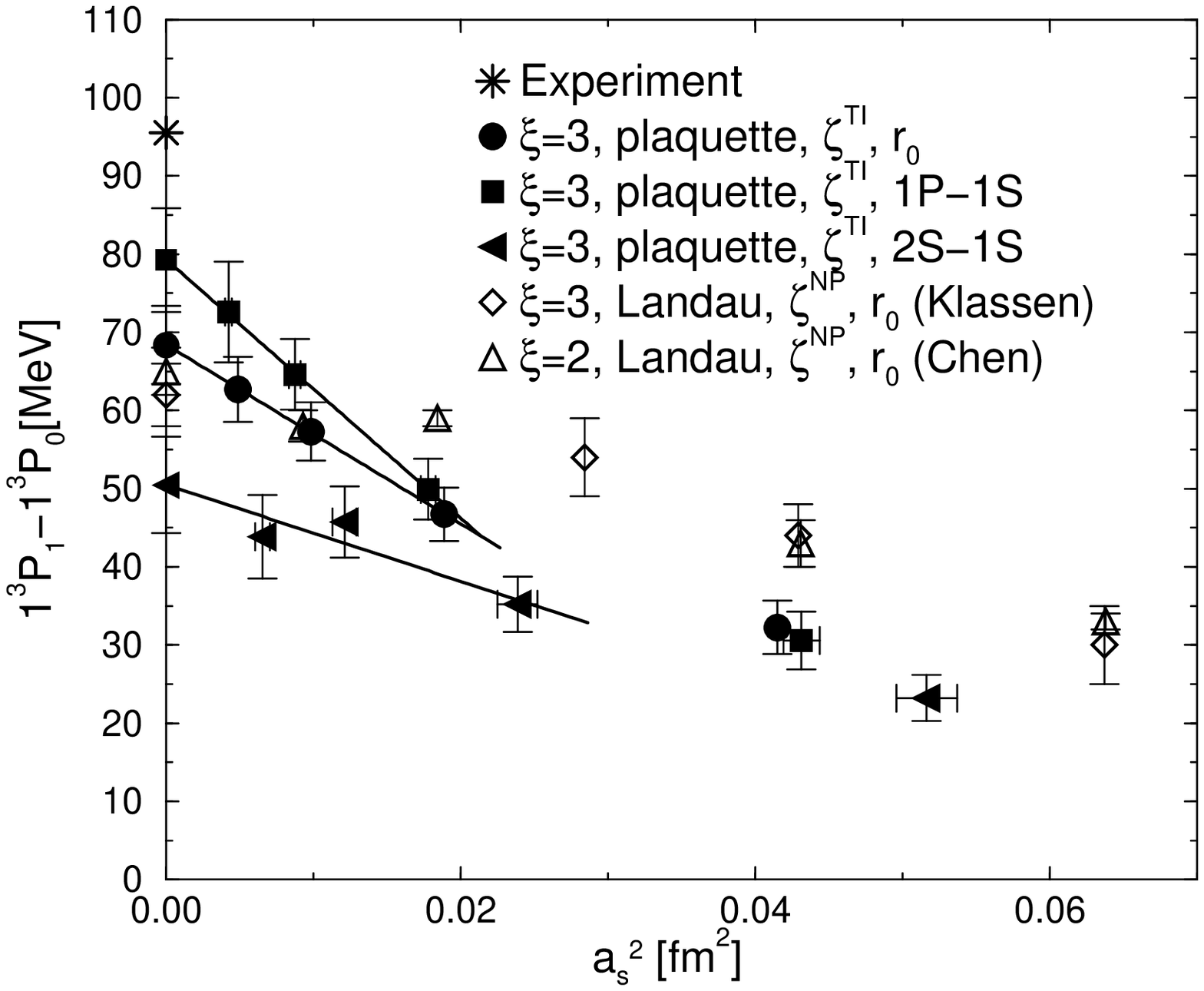}
\end{center}
\vspace{-16mm}
    \caption{Fine structure.}
    \label{fig:fss}
\vspace{-13pt}
\end{figure}

In Fig.\ref{fig:hfs}, we plot by filled symbols the lattice spacing 
dependence of the hyperfine splitting $\Delta M(1^3S_1-1^1S_0)$ 
for three inputs for the scale. 
Data at finite $a_s$ are extrapolated to the
continuum limit adopting an $a_s^2$-linear ansatz. 
The results largely depend on scale inputs, and are much smaller
than the experimental value ({\it e.g.,} by 
about 30\% with $1\bar{P} - 1\bar{S}$ input). 
Thus quenching effects are very large for the hyperfine splitting. 

In the same figure, we also plot results by Klassen 
(open diamonds; $\xi=3$)\cite{Klassen} and 
Chen (open triangles; $\xi=2$)\cite{Ping} with the same action. 
Their simulations differ from ours in that we determine the tadpole factor 
$u_0$ from the plaquette average and adopt for the parameter $\zeta$ 
the tree-level tadpole improved value $\zeta^{\rm TI}$, while they use 
the mean link in the Landau gauge for $u_0$
and a non-perturbative estimate $\zeta^{\rm NP}$
determined from the meson dispersion relation. 
Nonetheless, their results and ours, using the same scale $r_0$, 
all converge to a consistent value of about 70~MeV in the continuum limit.

\subsection{Fine structure}

Figure \ref{fig:fss} shows results of the fine structure 
$\Delta M(1^3P_1-1^3P_0)$. 
The deviation from the experimental value is smaller than
that for the hyperfine splitting 
(about 20\% with $1\bar{P} - 1\bar{S}$ input).
Our result with $r_0$ input is again 
consistent with those of Refs.~\cite{Klassen,Ping}.

\section{Effect of $c_s$ for hyperfine splitting} 

The results described so far all use the tadpole improved 
value $\tilde{c_s} = 1$ for the spatial clover coefficient. 
In Refs.\cite{Klassen98,Klassen}, Klassen employed a different 
choice $\tilde{c_s} = 1/\nu$ ($\nu \equiv \xi_0/\zeta$).  
He obtained ${\rm HFS}(a_s=0,r_0) \approx 90$~MeV for the continuum 
limit of the hyperfine splitting, which is much larger than the 
result above
${\rm HFS}(a_s=0,r_0) \approx 70$~MeV with $\tilde{c_s} = 1$.
We note that $\tilde{c_s} = 1/\nu$ is correct {\it only} in the massless 
limit, while $\tilde{c_s} = 1$ is valid for {\it any} quark mass, 
at the tree level.  

To resolve this problem, we attempt an effective analysis. 
The potential model predicts that the hyperfine splitting is due to the
spin-spin interaction of quarks, which originates from the 
${\bf\Sigma \cdot B}$ term 
in the nonrelativistic Hamiltonian $H^{\rm NR}$. 
We therefore define a ``tree-level effective hyperfine splitting'' 
\begin{equation}
{\rm HFS}^{\rm eff} \equiv 
( a_t\tilde{M}_{\rm 1} / a_t\tilde{M}_B )^2 \,\,\, ,
\label{effhfs}
\end{equation}
where 
\begin{equation}
\frac{1}{a_tM_B} = \frac{2\xi^2/\zeta^2}{m_0(2+m_0)}
+ \frac{\xi^2 {c_s}/\zeta}{1+m_0}
\end{equation}
is the tree level coefficient of the 
${\bf\Sigma \cdot B}$ term in $H^{\rm NR}$.
The pole mass $a_tM_1 = \log (1+m_0)$ is inserted to normalize 
to unity in the continuum limit, and tildes denote the tadpole improvement.

\begin{figure}[t]
\centerline{\epsfxsize=3.85cm \epsfbox{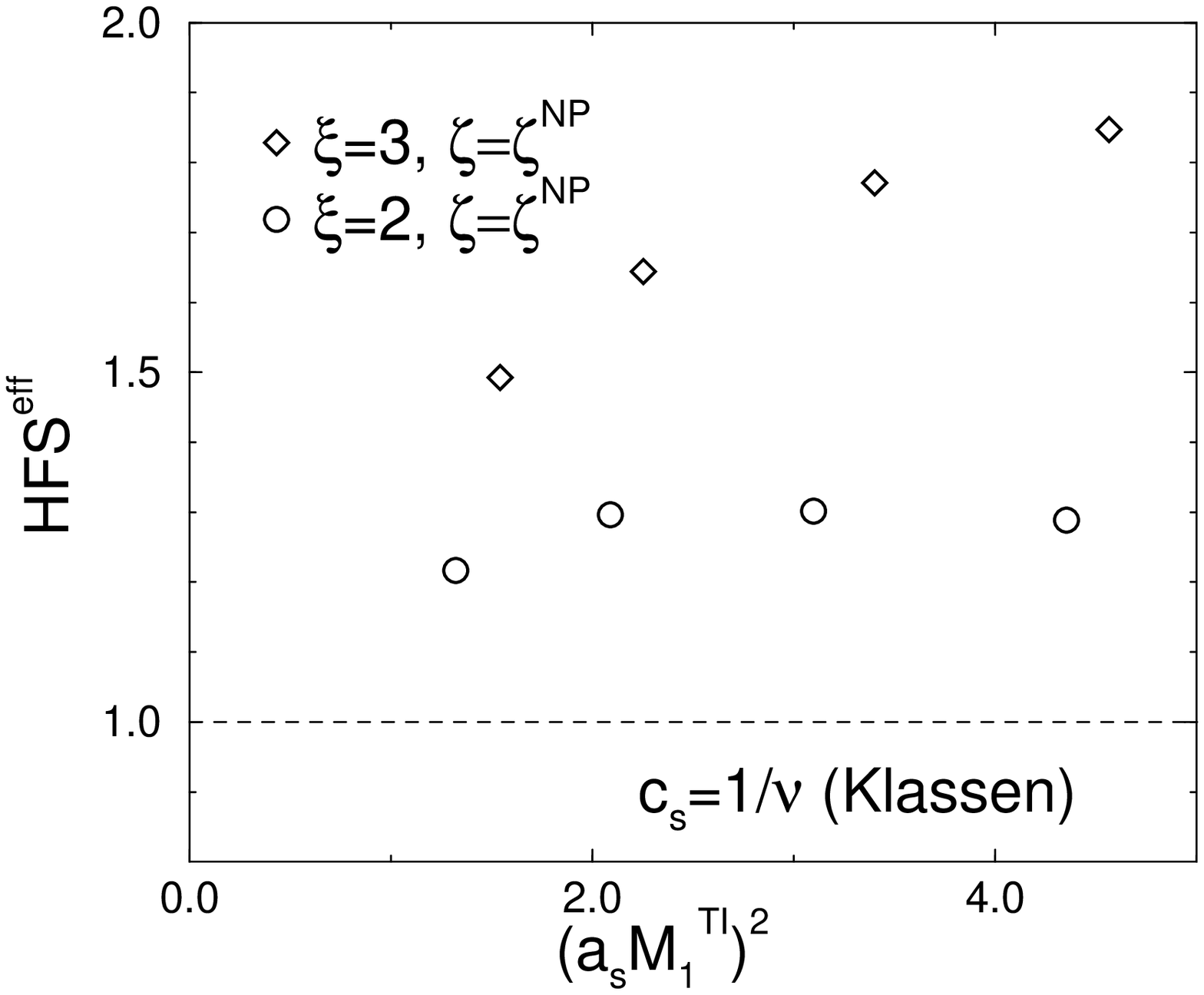}
\hspace{-0.cm}\epsfxsize=4.cm \epsfbox{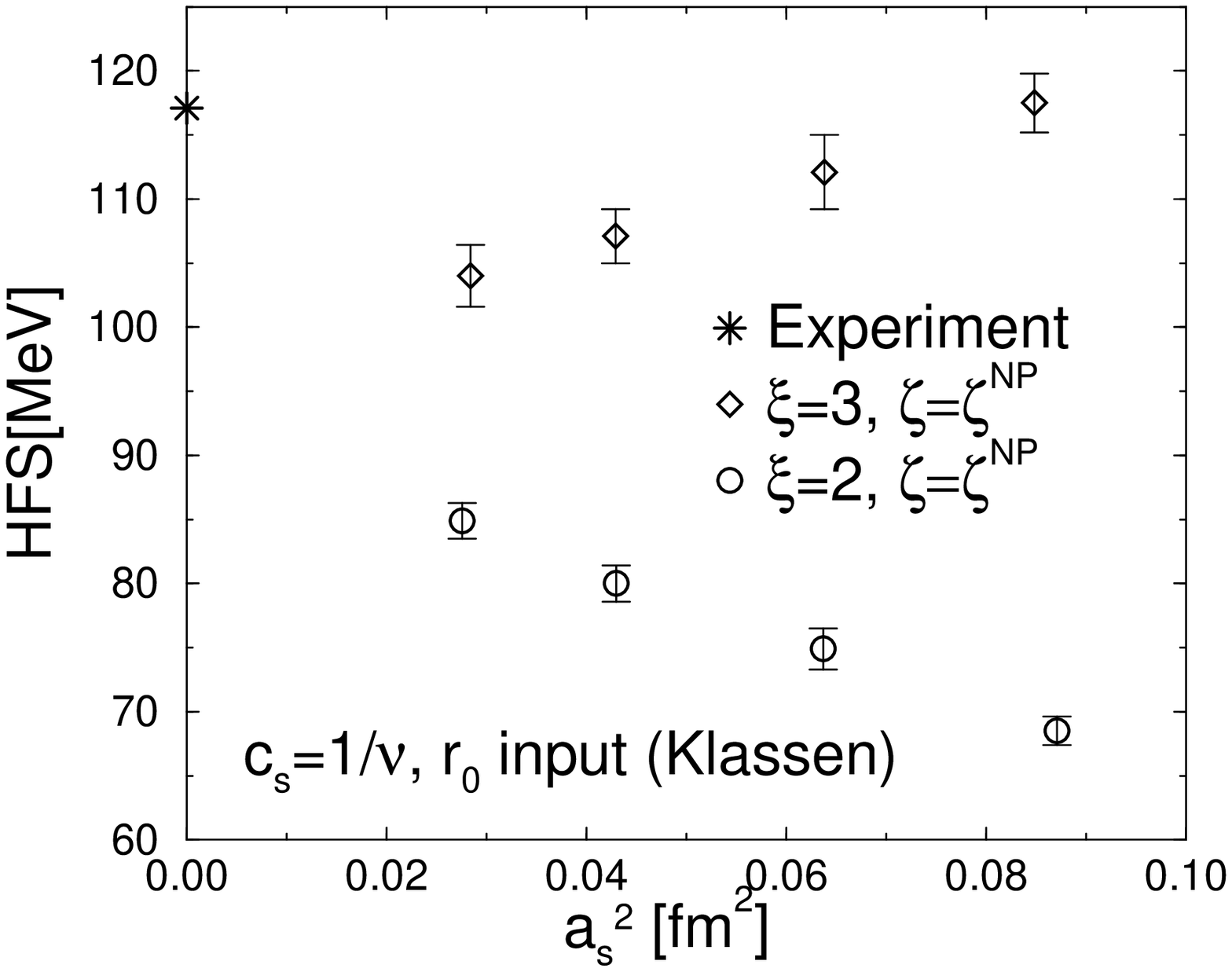}}
\vspace{-1cm}
\caption{${\rm HFS}^{\rm eff}$ and HFS for $\tilde{c_s}=1/\nu$.}
\label{fig:hfseffTK}
\vspace{0.3cm}
\centerline{\epsfxsize=3.85cm \epsfbox{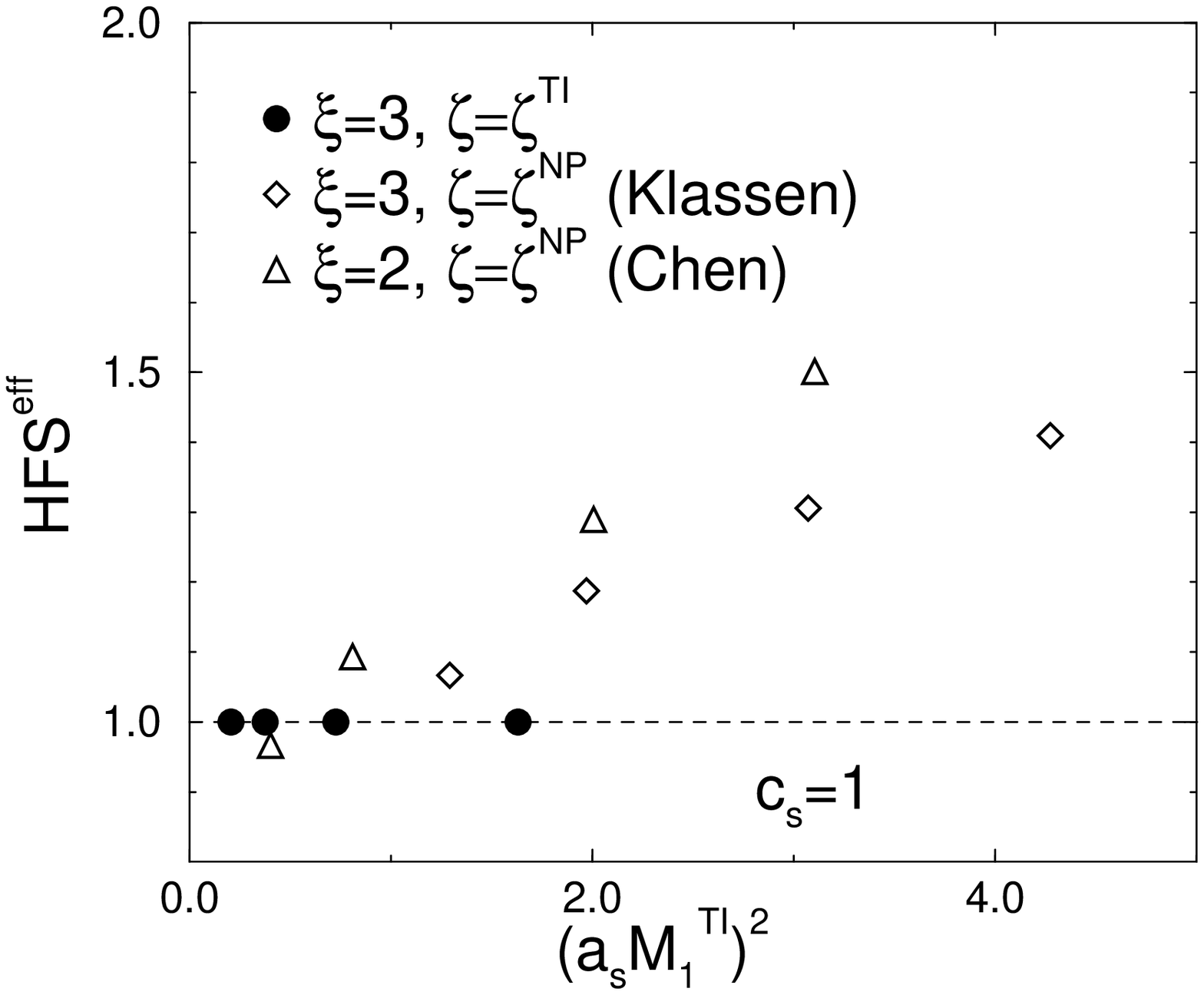}
\hspace{-0.cm}\epsfxsize=4.cm \epsfbox{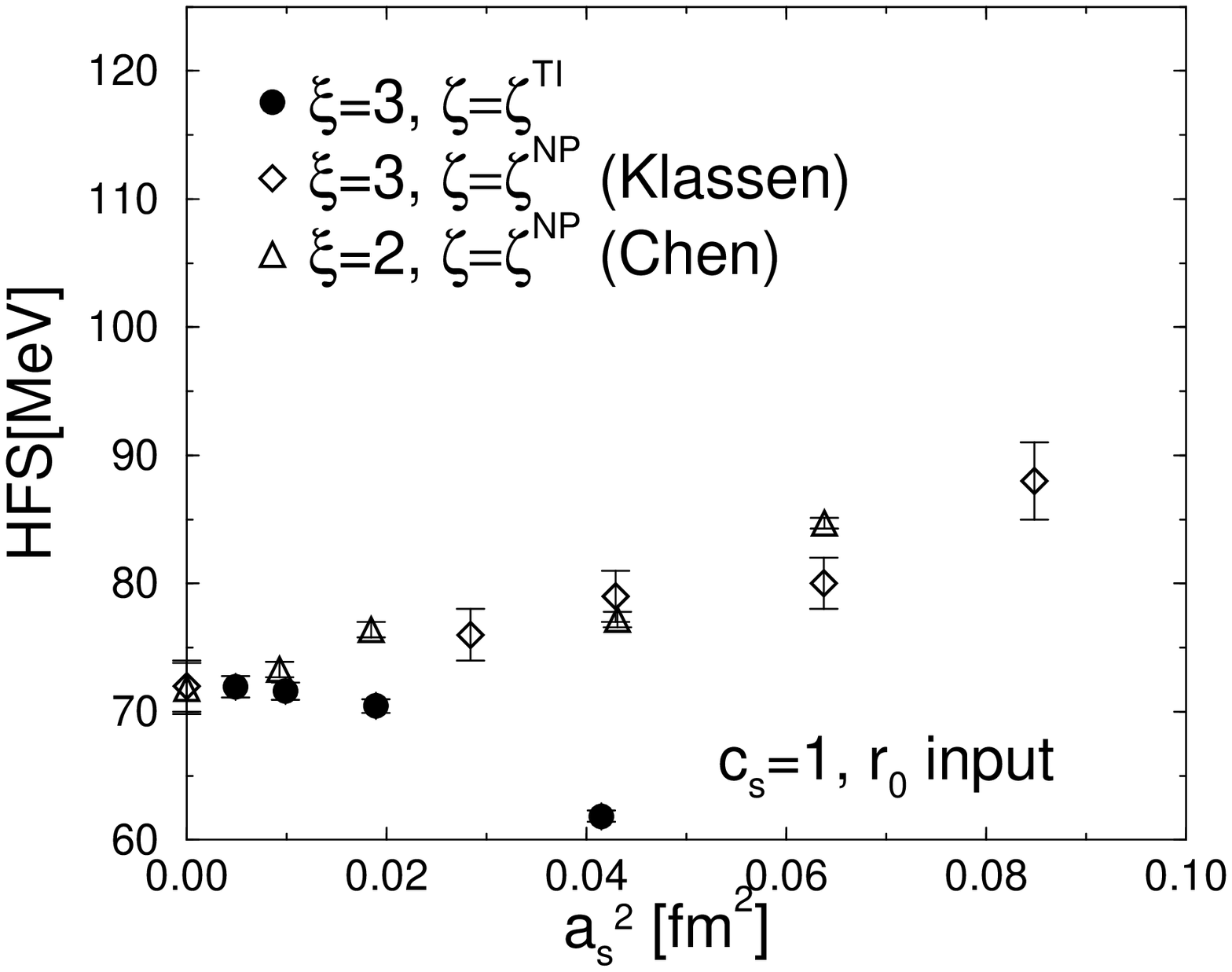}}
\vspace{-1cm}
\caption{${\rm HFS}^{\rm eff}$ and HFS for $\tilde{c_s}=1$.}
\label{fig:hfseffCP}
\vspace{-0.7cm}
\end{figure}

In Fig.\ref{fig:hfseffTK} we compare the scaling behavior of 
${\rm HFS}^{\rm eff}$ (left panel) 
and the actual data HFS (right panel) for $\tilde{c_s}=1/\nu$.  A similar 
comparison for $\tilde{c_s}=1$ is made in Fig.~\ref{fig:hfseffCP}. 
We find that results of HFS are qualitatively well reproduced by those of 
${\rm HFS}^{\rm eff}$.
For $\tilde{c_s} = 1/\nu$, ${\rm HFS}^{\rm eff}$ remains large even at 
$(a_s\tilde{M}_{\rm 1})^2 \sim 1$, which suggests that the actual 
HFS should rapidly decrease as $a_s \rightarrow 0$, and hence 
a naive estimation $\approx 90$~MeV\cite{Klassen98,Klassen}
from an $a_s^2$-linear continuum fit is misleading for this case.
On the other hand, ${\rm HFS}^{\rm eff}$ is already close to unity for 
$(a_s\tilde{M}_{\rm 1})^2 \lsim 1$ for $\tilde{c_s} = 1$.
Thus an $a_s^2$-linear continuum estimation ($\approx 70$~MeV) for this case 
appears much more reliable than that for $\tilde{c_s} = 1/\nu$. 
 
\section{Conclusions}

We have computed the charmonium  
spectrum accurately using quenched anisotropic lattices with $\xi=3$.  
We find that the spin splittings largely depend on the scale
input and are smaller than the experimental values.
Our results are consistent with previous results \cite{Klassen,Ping} 
when the same clover coefficients are used. 
We have also shown that a large hyperfine splitting reported 
in Ref.~\cite{Klassen98,Klassen} with a
different choice of the clover coefficients is likely an overestimate 
arising from the continuum extrapolation.

\vspace*{3mm}

This work is supported in part by Grants-in-Aid
of the Ministry of Education 
(Nos.~10640246, 
10640248, 
11640250, 
11640294, 
12014202, 
12304011, 
12640253, 
12740133, 
13640260).
VL is supported by JSPS Research for the Future Program
(No. JSPS-RFTF 97P01102).
KN and M. Okamoto 
are JSPS Research Fellows.

\end{document}